\documentclass[12pt,twoside]{article}
\usepackage{amsmath}

\input BoxedEPS

\SetRokickiEPSFSpecial  

\HideDisplacementBoxes

\newcommand{\bra}[1]{\left\langle #1 \right|}
\newcommand{\ket}[1]{\left|#1\right\rangle}

\begin{document}


\title{Each normalized state  is a member of an orthonormal basis: A simple proof}

\author{Iman Sargolzahi$^1$ and Ehsan Anjidani$^2$ \\
\small $^1$ Department of Physics, University of Neyshabur, P. O. Box 91136-899, Neyshabur, Iran.\\ sargolzahi@neyshabur.ac.ir\\
\small $^2$ Department of Mathematics, University
of Neyshabur, P. O. Box 91136-899, Neyshabur, Iran.\\
anjidani@neyshabur.ac.ir\\[-0.25in]}
\date{}
\maketitle

\pagestyle{myheadings}
\markboth{I. Sargolzahi and E. Anjidani}{Each normalized state  is a member of an orthonormal basis: A simple proof}
\thispagestyle{empty}

\begin{abstract}
\noindent	

In a finite dimensional Hilbert space, each normalized vector (state) can be chosen as a member of an orthonormal basis of the space. We give a proof of this statement in a manner that seems to be more comprehensible for physics students than the formal abstract one.
\end{abstract}
$$ $$

$ $
\section{Introduction}

Finite dimensional Hilbert spaces, e.g. spin of a particle or polarization of a photon, is of great interest in quantum mechanics, specially in quantum information and computation theory \cite{1}. So, studying the mathematical properties of such spaces is necessary for physics students, specially in graduate levels.

One useful result in a finite dimensional Hilbert space is that an arbitrary normalized state can be chosen as a member of an orthonormal basis of the space. In the linear algebra theory, this statement is proven \cite{2}: A set including only one state, constructs a linearly independent set. But this proof seems rather abstract, at least, for physics students. So in this paper, instead, we prove this statement by constructing directly a basis including our arbitrarily chosen state. Doing so, in a two dimensional Hilbert space (one-qubit state space), is simple using the Bloch sphere. Then, by induction, we generalize this result to the higher finite dimensional Hilbert spaces too. We hope that this method be more comprehensible for physics students.

The paper is organized as follows: In the next section, we review the properties of vector spaces briefly. In section 3, using the Bloch sphere, it is shown that in the one-qubit case, each state is a member of an orthonormal basis. Our main result is given in section 4, in which, by induction, we generalize the result of section 3 to arbitrary finite dimensional spaces. In section 5, we discuss the consequences of this result briefly. When our result can be used for the countable infinite dimensional case is the subject of section 6 and, finally, our paper is ended by a summary in section 7.

\section{Finite dimensional vector spaces}\label{s2}
A vector space $V$ on complex numbers $C$ is a set of elements, like $\ket a, \ket b, \ket c, \cdots$, which we call them vectors and satisfy two following properties \footnote {For more detailed definition of a vector space see e.g. \cite{2, 3, 4}} \cite{3}:
\begin{enumerate}
\item Sum of each two vectors $\ket a$ and $\ket b$ of $V$ is also a vector in $V$ like $\ket c$:
\begin{equation}
\ket a\in V,~ \ket b\in V \Rightarrow\ket a+\ket b=\ket c\in V.
\label{1}
\end{equation}
\item Multiplication of a vector $\ket a$ by a complex number (scalar) $z$, is also a member of $V$:
\begin{equation}
\ket a\in V,~ z\in C \Rightarrow z\ket a\in V.
\label{2}
\end{equation}
\end{enumerate}
Two above properties are obviously the generalization of the properties of real vectors in the three dimensional space:
\begin{enumerate}
\item Sum of two vectors $\textbf{A}$ and $\textbf{B}$ is also a vector like $\textbf{C}$.
\item Multiplication of a vector $\textbf{A}$ by a real scalar $c$ is also a vector (parallel to $\textbf{A}$).
\end{enumerate}
In addition, in a vector space there is a zero vector, which we denote by 0, with the property
\begin{equation}
 \ket a +0=\ket a ~~~~~~ {\rm {for ~ any~}} \ket a\in V.
 \label{3}
\end{equation}
By definition, a set of nonzero vectors $\{\ket {a_1},\cdots, \ket {a_n}\}$ are linearly independent if the equality
\begin{equation}
z_1\ket {a_1}+ \cdots+ z_n\ket {a_n}=0
\label{4}
\end{equation}
(with $z_i\in C$, $i=1, \cdots, n$) is satisfied only when for all $1\leq i\leq n$, $z_i=0$.
Assume that the maximum number of linearly independent vectors in the vector space $V$ is a finite number $n<\infty$.
Now, consider the following equality, in which the vector $\ket{b}$ is also nonzero:
\begin{equation}
z_1\ket {a_1}+ \cdots+ z_n\ket {a_n}+z\ket b=0.
\label{5}
\end{equation}
Obviously, the above equation has a trivial solution: $z_i=0$, ($i=1, \cdots, n$) and $z=0$. If there were no other solution except the trivial one, then $\{\ket{a_1},\cdots,\ket{a_n},\ket{b}\}$ would be a linearly independent set with $(n+1)$ members, which is in contradiction to our assumption. So, equation \eqref{5} has another solution; i.e.
at least one of $z_i$ and also $z$ are nonzero.
This has an important consequence: For each $\ket b\in V$, we have
\begin{eqnarray}
\nonumber \ket b &=& \frac{z_1}{-z}\ket{a_1}+\cdots+\frac{z_n}{-z}\ket{a_n} \\
 &=&  \sum_{i=1}^nc_i\ket{a_i},
 \label{6}
\end{eqnarray}
where $c_i:=\frac{z_i}{-z}$. In other words, any vector in the vector space $V$ can be written as a linear combination of vectors  $\ket{a_1}, \cdots, \ket{a_n}$.
The set $\{\ket{a_1},\cdots,\ket{a_n}\}$ is called a basis for the vector space $V$. Such a vector space, in which the number of the basis vectors is finite ($n<\infty)$, in other words, the maximum number of linearly independent vectors is finite, is called a finite dimensional ($n$-dimensional) vector space \cite{2,3}.

A finite dimensional vector space $V$ equipped with an inner product is called a Hilbert space or an inner product space \cite{1}. A function $(.,.)$ from $V\times V$ to $C$ is an inner product if it satisfies the following properties \cite{1,3,4}:
\begin{enumerate}
\item $(.,.)$ is linear in the second argument ($\lambda_i$ are complex numbers.):
\begin{equation}
\left(\ket a,\sum_{i=1}^m\lambda_i\ket{b_i}\right)=\sum_{i=1}^m\lambda_i(\ket a,\ket {b_i}).
\label{6'}
\end{equation}
\item For any $\ket a,\ket b\in V$,
\begin{equation}
(\ket{a},\ket{b})=(\ket{b},\ket{a})^*.
\label{7}
\end{equation}
\item For any $\ket a\in V$,
\begin{equation}
(\ket a,\ket a)\geq 0,
\label{8}
\end{equation}
where the equality occurs if and only if $\ket a=0$.
\end{enumerate}

We define the norm of a vector $\ket a$ by
\begin{equation}
\|\ket a\|=\sqrt{(\ket a,\ket a)} .
\label{10}
\end{equation}
If $\|\ket a\|=1$, then we say that $\ket a$ is a normalized vector. In addition, if
\begin{equation}
(\ket a,\ket b)=0,
\label{11}
\end{equation}
then we say that the two vectors $\ket a$ and $\ket b$ are orthogonal.

Following the Gram-Schmidt procedure \cite{3,4}, one can construct an orthonormal basis $\{\ket{e_1},\cdots,\ket{e_n}\}$ from the basis $\{\ket{a_1},\cdots,\ket{a_n}\}$ in an inner product space. The orthonormality of vectors $\ket{e_i}$ means that $(\ket{e_i},\ket{e_j})=\delta_{ij}$, where $\delta_{ij}$ is the Kronecker delta function ($\delta_{ij}=0$ ~for~ $i\neq j$ ~and~ $\delta_{ij}=1$ ~for~ $i=j$). The Gram-Schmidt procedure is as follows:
\begin{equation}
\ket{e_1}:=\frac{\ket{a_1}}{\|\ket{a_1}\|},
\label{12a}
\end{equation}
and for each $2\leq i\leq n$,
\begin{equation}
\ket{e_i}:=\frac{\ket{e_i'}}{\|\ket{e_i'}\|},
\label{12}
\end{equation}
where
\begin{equation}
\ket{e_i'}:=\ket{a_i}-\sum_{j=1}^{i-1}(\ket{e_j},\ket{a_i})\ket{e_j}.
\label{12'}
\end{equation}
It is easy to see that the set $\{\ket{e_1},\cdots,\ket{e_n}\}$ is orthonormal. In addition, since according to \eqref{12a}, \eqref{12} and \eqref{12'}  we can write each $\ket{a_i}$ in terms of vectors $\ket{e_j}$, for each $\ket{b}\in V$ we have
\begin{equation}
\ket b=\sum_{i=1}^nc_i\ket{a_i}=\sum_{i=1}^nd_i\ket{e_i},
\label{13}
\end{equation}
where the coefficients $d_i$ are complex numbers. We may use the column matrix notation for the orthonormal basis vectors $\ket{e_i}$:
\begin{equation}
\ket{e_1}=\left(
            \begin{array}{c}
              1 \\
              0 \\
              \vdots \\
              0 \\
            \end{array}
          \right)_{n\times 1}
          ,\quad\quad\cdots\quad\quad,
          \ket{e_n}=\left(
                      \begin{array}{c}
                        0 \\
                        \vdots \\
                        0 \\
                        1 \\
                      \end{array}
                    \right)_{n\times 1}.
 \label{14}
\end{equation}
Hence, for each $\ket b\in V$ we have
\begin{equation}
\ket b= \sum_{i=1}^n d_i\ket {e_i}=d_1\left(
                                        \begin{array}{c}
                                          1 \\
                                          0 \\
                                          \vdots \\
                                          0 \\
                                        \end{array}
                                      \right)
                                      +\cdots+ d_n\left(
                                                    \begin{array}{c}
                                                      0 \\
                                                      \vdots \\
                                                      0 \\
                                                      1 \\
                                                    \end{array}
                                                  \right)= \left(
                                                             \begin{array}{c}
                                                               d_1 \\
                                                               \vdots \\
                                                               d_n \\
                                                             \end{array}
                                                           \right).
 \label{15}
\end{equation}
Now, using the properties \eqref{6'} and \eqref{7} of the inner product function, it can be shown that for any $\ket v, \ket w\in V$, where
$$\ket v=\left(
           \begin{array}{c}
             v_1 \\
             \vdots \\
             v_n \\
           \end{array}
         \right)\quad\quad,\quad\quad\ket w=\left(
                                       \begin{array}{c}
                                         w_1 \\
                                         \vdots \\
                                         w_n \\
                                       \end{array}
                                     \right),$$
we have \cite{1}
\begin{equation}
(\ket v,\ket w)=\left(
                  \begin{array}{ccc}
                    v_1^* & \cdots & v_n^* \\
                  \end{array}
                \right)\left(
                         \begin{array}{c}
                           w_1 \\
                           \vdots \\
                           w_n \\
                         \end{array}
                       \right)=\sum_{i=1}^nv_i^*w_i.
\label{16}
\end{equation}
We can continue the above discussion and introduce linear operators on $V$, their matrix representation and ...  \cite{1,5}.

Let's come back to our question: whether an arbitrary normalized vector $\ket b\in V$ can be chosen as a member of an orthonormal basis. As we mentioned, the answer is yes, since the set $\{\ket b\}$ is a linearly independent set. Therefore $\ket b$ is the basis vector of a one dimensional subspace of  $V$ and so a basis vector of the $V$ itself \cite{2}. In addition, one can find exactly $n-1$ other nonzero vectors $\ket{w_i}$ in such a way that $\{\ket b, \ket{w_1},\cdots,\ket{w_{n-1}}\}$ constructs a linearly independent set and so a basis for $V$ (see \cite{2,3}). Equipped with the linearly independent set $\{\ket b, \ket{w_1},\cdots,\ket{w_{n-1}}\}$ and using the Gram-Schmidt procedure, we can construct the orthonormal basis $\{\ket{v_1}(=\ket b),\ket{v_2},\cdots,\ket{v_n}\}$ for $V$, and therefore the proof is completed.

The above discussion, though is satisfying for a mathematician, may be rather abstract for a physics student. So, instead, we prove that each normalized vector $\ket b$ is a member of an orthonormal basis, by constructing such a basis directly from the orthonormal computational basis $\{\ket{e_1},\cdots,\ket{e_n}\}$. This can be done simply for the two dimensional case, i.e. qubits, using the Bloch sphere. For the completeness of our discussion, we give this result in the next section and then in section 4, we generalize it to the higher dimensional cases.


\section{Two dimensional Hilbert space: one-qubit state space}


In a two dimensional Hilbert space, according to \eqref{13}, each (normalized) state can be written as:
\begin{equation}
\ket\psi=\alpha\ket 0+\beta\ket 1,
\label{17}
\end{equation}
where  we use the notation $\ket 0$ and $\ket 1$ for the orthonormal computational basis, instead of $\ket {e_1}$ and $\ket {e_2}$.
Decomposition coefficients $\alpha$ and $\beta$ are (in general) complex numbers. Since $\ket\psi$ is normalized, from \eqref{10} and \eqref{16} we have
\begin{equation}
|\alpha|^2+|\beta^2|=1.
\label{18}
\end{equation}
So, we can write \eqref{17} as follows
\begin{equation}
\ket \psi=e^{i\gamma}\left(\cos\frac{\theta}{2}\ket 0+e^{i\varphi}\sin\frac{\theta}{2}\ket 1\right),
\label{19}
\end{equation}
where $i=\sqrt{-1}$ and $\gamma,\theta,\varphi$ are real numbers. Comparing \eqref{17} and \eqref{19}, we have $$\alpha=e^{i\gamma}\cos\frac{\theta}{2}\quad,\quad\beta=e^{i(\gamma+\varphi)}\sin\frac{\theta}{2};$$
so \eqref{18} is obviously satisfied. Therefore, up to phase factor, we can write each $\ket\psi$ as \cite{1}:
\begin{equation}
\ket \psi=\cos\frac{\theta}{2}\ket 0+e^{i\varphi}\sin\frac{\theta}{2}\ket 1\quad\quad(0\leq\varphi\leq 2\pi).
\label{19a}
\end{equation}
Obviously, choosing $\varphi$ out of the above interval does not give us any new state. Also, it can be shown that choosing $0\leq \theta\leq \pi$ covers all possible states; i.e. if we choose $\theta$ out of this interval, then, up to a phase factor, it coincides with a state with $0\leq \theta\leq \pi$. \\ So, all possible one-qubit states can be written as follows:
\begin{equation}
\ket \psi=\cos\frac{\theta}{2}\ket 0+e^{i\varphi}\sin\frac{\theta}{2}\ket 1\quad\quad(0\leq\varphi\leq 2\pi,~~ 0\leq \theta\leq \pi).
\label{19b}
\end{equation}
Therefore, we can relate each state in a two dimensional Hilbert space to a point on a unit three dimensional sphere, called the Bloch sphere \cite{1}.

Now, consider an arbitrary state $\ket{\psi_0}$ (with~ $\theta=\theta_0$ ~and~ $\varphi=\varphi_0$):
\begin{equation}
\ket {\psi_0}=\cos\frac{\theta_0}{2}\ket 0+e^{i\varphi_0}\sin\frac{\theta_0}{2}\ket 1.
\label{20}
\end{equation}
It is easy to show that $\ket{\psi_0}$ is a member of an orthonormal basis for our one-qubit state space. Consider the state $\ket{\phi_0}$ (with~ $\theta=\pi-\theta_0$ ~and~ $\varphi=\varphi_0+\pi$):
\begin{equation}
\ket {\phi_0}=\cos\frac{\pi-\theta_0}{2}\ket 0+e^{i\varphi_0}\sin\frac{\pi-\theta_0}{2}\ket 1=\sin\frac{\theta_0}{2}\ket 0-e^{i\varphi_0}\cos\frac{\theta_0}{2}\ket 1.
\label{21}
\end{equation}
Using \eqref{16}, it is easy to show that $\ket {\psi_0}$ and $\ket{\phi_0}$ are orthonormal. In addition, we have
\begin{eqnarray}
\nonumber  \ket 0 &=& \cos\frac{\theta_0}{2}\ket {\psi_0}+\sin\frac{\theta_0}{2}\ket{\phi_0},\\
  \ket 1 &=& e^{-i\varphi_0}\left(\sin\frac{\theta_0}{2}\ket{\psi_0}- \cos\frac{\theta_0}{2}\ket{\phi_0}\right).
  \label{22}
\end{eqnarray}
So, we can decompose each state $\ket\psi$, in this two dimensional state space, in terms of $\ket{\psi_0}$ and $\ket{\phi_0}$. This can be done by inserting $\ket{0}$ and $\ket 1$ from \eqref{22} in \eqref{17}. Therefore, $\{\ket{\psi_0},\ket{\phi_0}\}$ is an orthonormal basis for one-qubit state space, which includes our arbitrarily chosen state $\ket{\psi_0}$.

\section{Higher dimensional Hilbert spaces}


Now we want to generalize the result of the previous section to an arbitrary finite dimensional Hilbert space. We begin with the three dimensional case: one-qutrit state space.

An arbitrary one-qutrit state can be decomposed in terms of the computational basis $\{\ket 0,\ket 1, \ket 2\}$ as follows
\begin{equation}
\ket \omega=\alpha'\ket 0+\beta'\ket 1+\gamma'\ket 2,
\label{23}
\end{equation}
where $\alpha',\beta',\gamma'$ are complex numbers. The normality condition is as follows
\begin{equation}
|\alpha'|^2+|\beta'|^2+|\gamma'|^2=1.
\label{23'}
\end{equation}

We want to show that the arbitrary state $\ket\omega$ is a member of an orthonormal basis. We do this by constructing such a basis explicitly. (We restrict ourselves to the case $\alpha',\beta',\gamma'\neq 0$, otherwise, the problem reduces to the lower dimensional cases.)
Consider the projection of $\ket \omega$ into the subspace spanned by $\{\ket 0,\ket 1\}$:
\begin{equation}
\ket{\psi'}=\alpha'\ket 0+\beta'\ket 1.
\label{24}
\end{equation}
So
\begin{equation}
\ket{\psi}=\frac{\alpha'}{\psi'}\ket 0+\frac{\beta'}{\psi'}\ket 1 :=\alpha\ket 0+\beta\ket 1\quad(\psi':=\|\ket{\psi'}\|)
\label{25}
\end{equation}
is a normalized vector in this two dimensional subspace. $\ket\psi$ is obviously in the form of \eqref{17}. So, up to phase factor, it can be written as \eqref{20} and we can find an orthogonal state to $\ket \psi$ as \eqref{21}:
\begin{equation}
\ket{\varphi}=\beta^*\ket 0-\alpha^*\ket 1.
\label{26}
\end{equation}
(Note that $\ket\varphi$ is also orthogonal to $\ket\omega$.)
Now, we can write \eqref{23} as:
\begin{equation}
\ket \omega=\psi'\ket \psi+\gamma'\ket 2,
\label{27}
\end{equation}
where  $\ket\psi$ and $\ket 2$ are orthonormal. So, $\ket\omega$ in \eqref{27} is also in the form of \eqref{17} which again can be written as \eqref{20} and we can find an orthogonal state to it as \eqref{21}:
\begin{equation}
\ket \nu=\gamma'^*\ket\psi-\psi'\ket 2;
\label{28}
\end{equation}
note that $\psi'=\|\ket{\psi'}\|$ is a positive number.
Now it is obvious that the set $\{\ket\varphi,\ket\omega,\ket\nu\}$ is an orthonormal set. Also an arbitrary state $$\ket a=\alpha_0\ket 0+\alpha_1\ket 1+\alpha_2\ket 2$$ in this one-qutrit state space, can be decomposed in terms of $\{\ket\varphi,\ket\omega,\ket\nu\}$: similar to \eqref{22}, we can write $\ket 0$ and $\ket 1$ in terms of $\ket\psi$ in \eqref{25} and $\ket\varphi$ in \eqref{26} and then we can write $\ket\psi$ and $\ket 2$ in terms of $\ket\omega$ in \eqref{27} and $\ket\nu$ in \eqref{28}. So, we can write $\ket a$ in terms of $\ket\varphi,\ket\omega$ and $\ket\nu$.

In summary, $\{\ket \varphi,\ket \omega,\ket \nu\}$, including our arbitrarily chosen state $\ket \omega$, constructs an orthonormal basis for three dimensional Hilbert space.

 Also note that $\ket\varphi$ and $\ket\nu$ span a two dimensional subspace, like $\ket 0$ and $\ket 1$ in \eqref{17}. So, similar to the previous section, instead of $\ket\varphi$ and $\ket\nu$, we can choose any other two orthonormal states $\ket{\varphi'}$ and $\ket{\nu'}$ which also span this subspace. Since $\ket{\varphi'}$ and $\ket{\nu'}$ are written in terms of $\ket\varphi$ and $\ket\nu$, the vectors $\ket{\varphi'}$ and $\ket{\nu'}$ are orthogonal to $\ket\omega$. Therefore, the set $\{\ket\omega,\ket{\varphi'},\ket{\nu'}\}$ is also an orthonormal basis for one-qutrit state space.

In other words, in a three dimensional Hilbert space (and also in higher dimensional cases) in contrast with the two dimensional one, we can find an infinite number of orthonormal bases including our chosen state $\ket\omega $.

In the two dimensional Hilbert space, for each $\ket{\psi_0}$ in \eqref{20}, up to a phase factor, there is only one $\ket{\phi_0}$ in \eqref{21} such that the set $\{\ket{\psi_0},\ket{\phi_0}\}$ constructs an orthonormal basis including $\ket{\psi_0}$.

Generalization of our result to the higher dimensions can be done by induction: Assume that in an $n$-dimensional Hilbert space, each arbitrary state is a member of an orthonormal basis. We want to show that it is also true for $(n+1)$-dimensional Hilbert space.

Let's show the computational basis as $\{\ket 1,\ket 2,\cdots\}$. So, in $(n+1)$-dimensional state space, we can decompose an arbitrary state $\ket\omega$ as
\begin{equation}
\ket\omega=\sum_{i=1}^{n+1}c_i'\ket i,
\label{29}
\end{equation}
where all $c_i'$ are complex numbers and the normality condition is as follows
\begin{equation}
\sum_{i=1}^{n+1}|c_i'|^2=1.
\label{29a}
\end{equation}
 We suppose that all $c_i'$ are nonzero, otherwise the problem reduces to the lower dimensional cases.

 Now, consider the projection of $\ket \omega$ into the $n$-dimensional subspace spanned by $\{\ket 1,\cdots,\ket n\}$:
 \begin{equation}
\ket{\psi'}=\sum_{i=1}^{n}c_i'\ket i.
\label{30}
\end{equation}
Hence
\begin{equation}
\ket\psi=\frac{\ket{\psi'}}{\psi'}=\sum_{i=1}^{n}\frac{c_i'}{\psi'}\ket i:=\sum_{i=1}^nc_i\ket i,\quad\quad(\psi':=\|\ket{\psi'}\|),
\label{31}
\end{equation}
is a normalized vector in this $n$-dimensional subspace. Since in the $n$-dimensional Hilbert space each state is a member of an orthonormal basis, we can find $n-1$ states $\ket{\varphi_i}$ such that $\{\ket{\varphi_1},\cdots,\ket{\varphi_{n-1}},\ket \psi\}$ constructs an orthonormal basis of the $n$-dimensional subspace spanned by $\{\ket 1,\cdots,\ket n\}$. Thus, $\{\ket{\varphi_1},\cdots,\ket{\varphi_{n-1}},\ket \psi,\ket{n+1}\}$ is an orthonormal basis of the $(n+1)$-dimensional Hilbert space.

In addition, we have
\begin{equation}
\ket\omega=\psi'\ket\psi+c_{n+1}'\ket{n+1}.
\label{32}
\end{equation}
States $\ket\psi$ and $\ket{n+1}$ span a two dimensional subspace which also can be spanned by orthonormal states $\ket\omega$ and
\begin{equation}
\ket\nu=c_{n+1}'^*\ket\psi-\psi'\ket{n+1}.
\label{33}
\end{equation}
Therefore the set $\{\ket{\varphi_1},\cdots,\ket{\varphi_{n-1}},\ket\omega,\ket\nu\}$, including the arbitrary state $\ket\omega$, is an orthonormal basis of $(n+1)$-dimensional state space and the proof is completed.

\section{Consequences}

In the previous section, we proved that in a finite dimensional state space, an arbitrary state $\ket\omega$ can be chosen as a member of an orthonormal basis. It seems valuable to discuss briefly the consequences of this result.

The first consequence of the above result is that, in addition to the computational basis, there exist other orthonormal bases for a finite dimensional state space. It has an important consequence: existence of unitary operators. A unitary operator $U$ is an operator  satisfying the following relations \cite{5}:
\begin{equation}
UU^\dag=I\quad\quad,\quad\quad U^\dag U=I,
\label{34}
\end{equation}
where $U^\dag$ is the adjoint of $U$ and $I$ is the identity operator. Now if we consider two orthonormal bases $\{\ket 1,\cdots,\ket n\}$ and $\{\ket{\varphi_1},\cdots,\ket{\varphi_n}\}$ for an $n$-dimensional Hilbert space, then it can be shown simply that
\begin{equation}
U=\sum_{i=1}^n\ket{\varphi_i}\bra i
\label{35}
\end{equation}
is a unitary operator \cite{5}.

Also, using this fact that each state can be chosen as a member of an orthonormal basis, one can prove the Cauchy-Schwarz inequality, as given in \cite{1}.

\section{Countable infinite dimensional Hilbert spaces}

Till now, our discussion was restricted to the finite dimensional Hilbert spaces, which are used widely in quantum information and computation theory \cite{1}. However, countable infinite dimensional Hilbert spaces are also of interest in quantum theory. Standard examples in this context are the harmonic oscillator and the infinite well \cite{6}.

Proving, in the general case, that an arbitrary normalized state $\ket \omega$, in a countable infinite dimensional Hilbert space $V$, is a member of an orthonormal basis of $V$, is out of the scope of this paper. But, there is an important special case for which we can prove the above statement simply, using our result in section 4.

Assume that the set $\{\ket 1,\ket 2,\cdots\}$ is the orthonormal computational basis of our countable infinite dimensional Hilbert space $V$. So we can write each $ \ket\psi \in V$ as
\begin{equation}
\ket\psi=\sum_{i=1}^{\infty} c_i\ket i,
\label{36}
\end{equation}
where $c_i$ are complex numbers.
Now, consider the special case that our arbitrarily chosen state $ \ket\omega$ can be written as
\begin{equation}
\ket\omega=\sum_{i=1}^{N} \alpha_{i}\ket i ,
\label{37}
\end{equation}
where $N$ is a finite positive integer. Therefore,  $\ket \omega$ belongs to a finite dimensional subspace of $V$, spanned by  $\{\ket 1,\ket 2,\cdots, \ket N\}$. Let's denote this subspace as  $V^{\prime}$. Now, using the result of section 4, we can find an orthonormal basis for $V^{\prime}$ which includes $\ket \omega$: $\{\ket \omega =\ket {e_{1}}, \ket {e_{2}},\cdots , \ket{ e_{N}}\}$. So,  the whole $V$  can be spanned by the orthonormal basis  $\{\ket \omega =\ket {e_{1}}, \ket {e_{2}},\cdots , \ket{ e_{N}}, \ket {N+1}, \ket {N+2} ,\cdots \}$, which includes our chosen state $\ket \omega$.

\section{Summary}

In a finite dimensional Hilbert space, each normalized state can be chosen as a member of an orthonormal basis of the space. Instead of the formal proof of this statement, we give a simple proof by constructing such a basis directly. In two dimensional case, this can be done by using the Bloch sphere. We generalize the two dimensional case to the higher dimensional state spaces by induction. We hope that our proof be more clear for physics students than the formal one.

In addition, even when the Hilbert space is countable infinite dimensional, we can use the above result, for an important special case, given in section 6.

\subsection*{Acknowledgments}

We would like to thank an anonymous referee for his (her) suggestion to add a section about the infinite dimensional case.

%


\begin{thebibliography} {9}

\bibitem{1} M. A. Nielsen and I. L. Chuang,   \emph{Quantum Computation and Quantum Information} (Cambridge University Press, 2000).

\bibitem{2} K. Hoffman and R. Kunze,  \emph{Linear Algebra} (Prentice-Hall, 1971) Chap. 2.

\bibitem{3} S. Hassani,  \emph{Foundations of Mathematical Physics} (McGraw-Hill, 1991) Chap. 2.

\bibitem{4} A. W. Joshi,  \emph{Matrices and Tensors in Physics} (John Wiley and Sons, 1995) Chap. 1.

\bibitem{5} J. J. Sakurai and J. Napolitano,   \emph{Modern Quantum Mechanics} (Addison-Wesley, 2011) Chap. 1.

\bibitem{6} S. Gasiorowicz, \emph{Quantum Physics} (John Wiley and Sons, 2003).

\end{thebibliography}
\end{document}